\documentclass[showpacs,preprintnumbers,amsmath,amssymb,twocolumn]{revtex4}

\usepackage{amsthm}
\usepackage{enumerate}

 \newtheorem{theorem}{Theorem}
 \newtheorem{lemma}[theorem]{Lemma}
 
 \newtheorem{definition}[theorem]{Definition}

\newcommand*{\bbN}{\mathbb{N}}

\newcommand*{\bbC}{\mathbb{C}}

\newcommand*{\cR}{\mathcal{R}}

\newcommand*{\cS}{\mathcal{S}}

\newcommand*{\cV}{\mathcal{V}}

\newcommand*{\cX}{\mathcal{X}}
\newcommand*{\cY}{\mathcal{Y}}

\newcommand*{\measp}{\mathcal{M}}
\newcommand*{\meas}{\mathcal{N}}

\newcommand*{\eps}{\varepsilon}
\DeclareMathOperator*{\tr}{tr}
\newcommand*{\id}{\mathrm{id}}
\DeclareMathOperator*{\ExpE}{E}

\newcommand*{\ket}[1]{| #1 \rangle}
\newcommand*{\bra}[1]{\langle #1 |}
\newcommand*{\proj}[1]{\ket{#1}\bra{#1}}

\newcommand*{\dist}[2]{\| #1 - #2 \|}
\newcommand*{\distb}[2]{\bigl\| #1 - #2 \bigr\|}

\newcommand*{\mysigma}[2]{\sigma_{#1#2}} 

\newcommand*{\cqstate}{$\{cq\}$-state}
\newcommand*{\ccqstate}{$\{ccq\}$-state}

\newcommand*{\nZ}{Z}
\newcommand*{\nz}{z}

\newcommand*{\nV}{V}
\newcommand*{\nv}{v}
\newcommand*{\ncV}{\cV}
\newcommand*{\nK}{X}
\newcommand*{\ncK}{\cX}
\newcommand*{\nk}{x}
\newcommand*{\nT}{Y}
\newcommand*{\ncT}{\cY}
\newcommand*{\nt}{y}
\newcommand*{\nE}{E}

\newcommand*{\nVE}{{\nV\nE}}

\newcommand*{\Iacc}{I_{\mathrm{acc}}}

\begin{document}

\title{Locking of accessible information and implications for the
  security of quantum cryptography}

\author{Robert K\"onig}

\email{r.t.koenig@damtp.cam.ac.uk}

\author{Renato Renner}

\email{r.renner@damtp.cam.ac.uk}

\affiliation{Centre for Quantum
   Computation \\ University of Cambridge \\ United Kingdom}

\author{Andor Bariska}
 
\email{andor.m.bariska@weiss.ch}


\author{Ueli Maurer}


\email{maurer@inf.ethz.ch}

\affiliation{Institute of Theoretical Computer Science \\ ETH Zurich \\ Switzerland}



\begin{abstract}
  
  

  
  The unconditional security of a quantum key distribution protocol is
  often defined in terms of the \emph{accessible information}, that
  is, the maximum mutual information between the distributed key $S$
  and the outcome of an optimal measurement on the adversary's
  (quantum) system.  We show that, even if this quantity is small,
  certain parts of the key $S$ might still be completely insecure when
  $S$ is used in applications, such as for one-time pad encryption.
  This flaw is due to a \emph{locking} property of the accessible
  information: one additional (physical) bit of information might
  increase the accessible information by more than one bit.

\end{abstract}

\maketitle

\section{Secrecy in classical and quantum cryptography} \label{sec:intro}

Secret keys play an important role in cryptography. They are used for
various tasks such as the encryption or authentication of messages.
Clearly, the security of these cryptographic tasks strongly depends on
the \emph{level of secrecy} of the underlying key.

The strongest and thus most desirable notion of security for a secret
key $S$ is called \emph{perfect security} and is characterized by two
conditions:
\begin{enumerate}[(i)]
\item any value of $S$ is equally likely (i.e., the distribution $P_S$
  is uniform on a \emph{key space~$\cS$});
  \label{cond:one}
\item an adversary has no information on~$S$ (i.e., the state of any
  system controlled by an adversary is independent of the value
  of~$S$).
  \label{cond:two}
\end{enumerate}
Such a perfectly secure key allows for the realization of highly
secure cryptographic schemes. For example, if $S$ is used as a
one-time pad~\footnote{In a \emph{one-time pad} encryption
  scheme~\cite{Vernam26}, a string $M$ of message bits is encrypted
  with a key $S$ of the same length. The ciphertext $C$ is given by
  bitwise addition (modulo $2$) of $M$ and $S$.} to encrypt a message
$M$, the resulting ciphertext $C$ is independent of $M$ and thus
completely useless for an adversary.

It turns out, however, that---even with the help of quantum
mechanics---it is generally impossible to generate perfectly secure
keys. One thus usually considers slightly weakened security
definitions. For example, condition~(ii) might be substituted by a
bound on the information that the adversary has on $S$.  This,
however, raises questions such as: What is an appropriate measure to
quantify the adversary's information on~$S$?  How to choose the upper
bound on this information such that it is guaranteed that $S$ can
safely be used in applications?

In the context of \emph{classical information-theoretic
  cryptography}~\footnote{For an introduction to classical
  information-theoretic key agreement, see, e.g., \cite{Maurer93}.},
the adversary's knowledge on a key $S$ is most generally characterised
by a classical random variable $Z$.  An $n$-bit key $S$ is then said
to be secure~\footnote{See, e.g.,
  \cite{Wyner75,BeBrRo88,CsiKor78,Maurer93}.} if, for some small $\eps
\geq 0$,
\begin{align}
  H(S) & \geq n-\eps \label{eq:HS}\\
  I(S;Z) & \leq \eps \label{eq:ISZ}
\end{align}
where $H(S)$ denotes the \emph{Shannon entropy} of $S$ and $I(S;Z) :=
H(S) - H(S|Z)$ is the \emph{mutual information between} $S$ and $Z$.
Inequality~\eqref{eq:HS} implies that $S$ is almost uniformly
distributed; it is thus an approximation of condition~(\ref{cond:one})
above.  Similarly, \eqref{eq:ISZ} is an approximation
of~(\ref{cond:two}).

In \emph{quantum cryptography} the knowledge of an adversary on a
(classical) key $S$ is described by the state of a quantum system $E$
instead of a classical random variable~$Z$.  Accordingly, the mutual
information occurring in criterion~\eqref{eq:ISZ} is thus usually
generalised to the \emph{accessible information} $\Iacc(S;E)$, which
is defined as the mutual information between $S$ and the outcome $Z$
of an optimal measurement applied to $E$ (see
Section~\ref{sec:locking} for a formal definition). The quantum
version of~\eqref{eq:ISZ} then reads
\begin{equation} \tag{\ref{eq:ISZ}$'$} \label{eq:Iacc} 
  \Iacc(S;E) \leq \eps \ .
\end{equation}
Inequality~\eqref{eq:Iacc} seems to be a natural formalisation of the
requirement that an adversary has almost no information on $S$ and is
in fact commonly used in the standard literature on quantum
cryptography and, in particular, quantum key
distribution~\footnote{See, e.g.,
  \cite{LoCha99,ShoPre00,NieChu00,GotLo03,LoChAr05} and also the
  discussion in~\cite{BHLMO05} and~\cite{RenKoe05}.}. However, as we
shall see, it is generally not sufficient to guarantee secrecy.

The remaining part of the paper is organized as follows. In
Section~\ref{sec:locking}, we review the definition of accessible
information and its locking property.  Section~\ref{sec:example} is
devoted to an explicit example of locking of the accessible
information.  This example is then used in Section~\ref{sec:problem}
to show that, even if the accessible information of an adversary on
the key~$S$ is arbitrarily small, $S$ might still be insecure for
certain applications.  Finally, in Section~\ref{sec:alternative}, we
discuss an alternative security definition which overcomes this
problem.

\section{Locking of accessible information} \label{sec:locking}

Let $\nE$ be a quantum system whose state depends on the value of a
classical random variable~$\nV$. This situation may be described using
the so-called \emph{enlarged Hilbert space representation} by encoding
the random variable $\nV$ into a quantum system with respect to an
orthonormal basis $\{\ket{\nv}\}_{\nv\in\ncV}$ as follows:
\begin{align*}
\rho_{\nVE}:=\sum_{\nv\in\ncV}P_\nV(\nv)\ \ket{\nv}\bra{\nv}\otimes\rho_{\nE|\nV=\nv}  \ ,
\end{align*}
where $\rho_{\nE|\nV=\nv}$ is the state of $\nE$ conditioned on $\nV =
\nv$.  We will refer to a state of this form as a \emph{\cqstate}. We
will also use generalisations of this convention to triparite systems
with two classical parts and call the corresponding states
\emph{\ccqstate{}s}.



For any \cqstate{} $\rho_{\nV \nE}$, the \emph{accessible information}
(of $\nE$ on $\nV$) is defined as~\footnote{In the literature, the
  accessible information is often defined in terms of ensembles. It is
  easy to verify that such a definition is equivalent to the one given
  here.}
\[
  \Iacc(\nV;\nE):=\max_{\measp} I(\nV;\nZ)
\]
where the maximum is over all local POVMs $\measp$ on $\nE$ and where
$I(\nV;\nZ)$ denotes the mutual information between $\nV$ and the
measurement outcome $\nZ$. The accessible information $\Iacc(\nV;\nE)$
thus quantifies the amount of information on the classical value $\nV$
that can be obtained by an optimal measurement applied to the quantum
system~$\nE$.


Consider now an extended setting involving an additional random
variable $\nT$, that is, the situation is described by a \ccqstate{}
$\rho_{\nV \nT \nE}$. Let~\footnote{$\Iacc(\nV;\nT \nE)$ denotes the
  accessible information of the \cqstate{} $\rho_{\nV(\nT\nE)}$ which
  is obtained from $\rho_{\nV \nT \nE}$ by combining the systems $\nT$
  and $\nE$.}
\[
\Delta := \Iacc(\nV; \nT \nE) - \Iacc(\nV;\nE)
\]
be the amount by which the accessible information on $\nV$ increases
when $\nT$ is appended to $\nE$. The quantity~$\Delta$ thus measures
by how much the knowledge on $\nV$ increases if one learns $\nT$
(given access to the quantum system $\nE$).
Interestingly, $\Delta$ can generally be larger than the \emph{size}
of $\nT$, i.e., the number of bits which are needed to represent its
value.  This phenomenon is known as \emph{locking}~\cite{DHLST04} and
will be the main topic of the next section.  It should be emphasized
that locking is a purely non-classical property. In fact, if the
quantum system $\nE$ is substituted by a classical random variable
$\nZ$, we have $\Delta = I(\nV;\nT|\nZ) \leq
H(\nT)$~\footnote{$I(\nV;\nT|\nZ) := H(\nV|\nZ) + H(\nT|\nZ) - H(\nV
  \nT| \nZ)$ is the mutual information between $\nV$ and $\nT$
  given~$\nZ$.}, that is, $\Delta$ cannot be larger than the size of
$\nT$.

\section{An example of locking} \label{sec:example}

\newcommand*{\sigmax}{\sigma_x}
\newcommand*{\sigmay}{\sigma_y}
\newcommand*{\sigmaz}{\sigma_z}
\newcommand*{\ntb}{\nt}
\newcommand*{\nTb}{\nT}
\newcommand*{\ntbp}{\nt'}

In this section, we give an explicit example of locking.  Compared to
previously known constructions~\cite{DHLST04,HLSW04,BCHLW05}, it has
some additional properties which are needed for our considerations
related to cryptography (see Section~\ref{sec:problem}).

In order to formulate our example of locking, we use the following
notational conventions: $\sigma_1,\sigma_2,\sigma_3$ are the Pauli
matrices on the Hilbert space $\bbC^2$. For any $m$-tuple $\ntb =
(\nt_1, \ldots, \nt_m)$ on $\{1, 2, 3\}$, we denote by $\sigma_{\ntb}$
the $m$-fold tensor product $\sigma_{\nt_1} \otimes \cdots \otimes
\sigma_{\nt_m}$.  Lemma~\ref{lem:propertypauli} summarises some
properties of these operators, which we will use repeatedly in the
following.

\newcommand*{\rhocE}[1]{\rho_{\nE|(\nK,\nTb)=(#1,\ntb)} }

Let $\nK$ and $\nT$ be random variables on the binary set $\ncK :=
\{0,1\}$ and the set of $m$-tuples $\ncT := \{1,2,3\}^m$,
respectively, such that the joint probability distribution $P_{\nK
  \nT}$ is uniform.  Moreover, for any $\nk \in \ncK$ and $\ntb \in
\ncT$, let
\begin{equation} \label{eq:operatordef}
\rhocE{\nk}
:= 2^{-m} \bigl(\id_{(\bbC^2)^{\otimes m}}
+ (-1)^x \sigma_{\ntb} \bigr)
\end{equation}
be an operator on $(\bbC^2)^{\otimes m}$, representing the state of a
quantum system $\nE$ conditioned on $\nK=\nk$ and $\nT = \nt$.
It is straightforward to check that this is a consistent description
of a \ccqstate{} $\rho_{\nK \nT \nE}$~\footnote{An alternative
  description of the state $\rho_{XYE}$ which clarifies the relation
  to the locking construction of \cite{DHLST04} is the following.  For
  $y\in\{1,2,3\}$, let $\{[0]_y,[1]_y\}$ denote the projectors onto
  the eigenspaces of $\sigma_y$. Let $R$ and $Y$ be independent and
  uniformly distributed random variables on $\cR:=\{0,1\}^m$ and
  $\cY$, respectively. For $r\in \cR$ and $y\in\cY$, let \[
  \rho_{E|(R,Y)=(r,y)}:=[r_1]_{y_1}\otimes \cdots \otimes [r_m]_{y_m}\ 
  . \] Finally, let $X$ be the random variable on $\cX$ defined by
  $X:=\oplus_{i=1} ^m R_i$, where $\oplus$ denotes addition
  modulo~$2$. It is then straightforward to check that the resulting
  conditional states $\rho_{E|(X,Y)=(x,y)}$ are given
  by~\eqref{eq:operatordef}.}.

Note that for any fixed $\nt \in \ncT$, the conditional quantum states
$\rhocE{0}$ and $\rhocE{1}$ are orthogonal. In particular, given
access to the quantum system $\nE$, the value of $\nK$ can be
determined with certainty if $\nT$ is known, that is, we have the
following statement.

\begin{lemma} \label{lem:certainty}
  Let $\rho_{\nK \nT \nE}$ be the \ccqstate{} defined above. For any
  fixed value $\nt \in \ncT$ of the random variable $\nT$, there
  exists a measurement of the quantum system $\nE$ with output equal
  to $\nK$.
\end{lemma}

On the other hand, if the value of $\nT$ is unknown, then any
measurement on $\nE$ reveals almost no information on the pair $(\nK,
\nT)$.

\begin{lemma}\label{lem:mainlockinglemma}
 Let $\rho_{\nK \nT \nE}$ be the \ccqstate{} defined above. Then
$ \Iacc(\nK \nT;\nE) \leq \bigl(\frac{2}{3}\bigr)^\frac{m}{2} $.
\end{lemma}

\begin{proof}  
  We show that, for any measurement $\measp$ applied to the quantum
  part $\nE$ of $\rho_{\nK \nT \nE}$ with outcome $\nZ$, the entropy
  of the pair $(\nK, \nT)$ conditioned on $\nZ$ is bounded by
  \begin{equation} \label{eq:goal}
    H(\nK \nT | \nZ) \geq H(\nK \nT) -  \textstyle \bigl(\frac{2}{3}\bigr)^\frac{m}{2} \ .
  \end{equation}
  The assertion then follows because $\Iacc(\nK \nT;\nE) = H(\nK \nT)
  - \min_{\measp} H(\nK \nT|\nZ)$.
  
  Let $\meas := \{d \cdot P_{\nK \nT}(\nk,\nt)\cdot \rho_{\nE|(\nK,
    \nT)=(\nk, \nt)}\}_{(\nk, \nt) \in \ncK \times \ncT}$ where
  $d:=2^m$ is the dimension of $\nE$.  Because $\rho_{\nE}$ is the
  fully mixed state on $\nE$, $\meas$ is a POVM on $\nE$.  By a
  similar derivation as in~\cite{DHLST04}, it can be shown that
  \begin{equation} \label{eq:Hbound}
    H(\nK \nT | \nZ) \geq \min_{\sigma} H(\meas[\sigma])
  \end{equation}
  where the minimum ranges over all states $\sigma$ on $\nE$ and
  $H(\meas[\sigma])$ is the entropy of the outcome when the
  measurement $\meas$ is applied to $\sigma$ (see
  Lemma~\ref{lem:accesslower} in the Appendix).
  
Using the fact that $\rho_{\nT \nE} = \rho_{\nT} \otimes \rho_{\nE}$
where $\rho_{\nE}$ is the fully mixed state, the term in the minimum
of~\eqref{eq:Hbound} can be rewritten as~\footnote{$E_{\nt \leftarrow
    P_{\nT}}[\cdot]$ denotes the expectation over the values $\nt$
  chosen according to the distribution $P_{\nT}$.}
  \begin{align}\label{eq:xyeentrop}
    H(\meas[\sigma])
  =
    H(\nT)+ \ExpE_{\nt\leftarrow P_\nT} [H(\meas_{\nt}[\sigma])]
\end{align}
where, for any $\nt \in \ncT$, $H(\meas_{\nt}[\sigma])$ is the entropy
of the output of the POVM $\meas_{\nt} := \{ d \cdot
P_{\nK|\nT=\nt}(\nk)\cdot \rho_{\nE|(\nK, \nT)=(\nk,
  \nt)}\}_{\nk\in\ncK}$ applied to $\sigma$.  Because for every $\nt
\in \ncT$ the POVM $\meas_{\nt}$ is binary-valued, this quantity is
easy to bound.  More precisely, as the binary entropy function
$h(p):=-p\log p-(1-p)\log (1-p)$ satisfies $h(p)\geq 1-|p-(1-p)|$ for
every $p\in [0,1]$ and
\[
|\tr(\sigma
(\rhocE{0}-\rhocE{1}))|=2^{-m+1}\tr(\sigma_y\sigma)
\]
for every $y\in\cY$ and every state $\sigma$ on
$(\mathbb{C}^2)^{\otimes m}$, we obtain by a straightforward
calculation
\begin{align}\label{eq:dervtwo}
 \ExpE_{\nt\leftarrow P_\nT} [H(\meas_{\nt}[\sigma])]
\geq 
  1-\frac{1}{|\ncT|}\sum_{\ntb\in\ncT} \bigl| \tr(\sigma_y\sigma) \bigr| \ .
\end{align}
Applying the Cauchy-Schwarz inequality gives
\begin{align}\label{eq:derivatone}
\frac{1}{|\ncT|}
\sum_{\ntb\in\ncT} \bigl| \tr(\sigma_y\sigma) \bigr| &\leq
\frac{1}{\sqrt{|\ncT|}}\sqrt{\sum_{\ntb\in\ncT}
\tr(\mysigma{\ntb}{}\sigma)^2}\leq \biggl(\frac{2}{3}\biggr)^{\frac{m}{2}}\ ,
\end{align}
where the last inequality is a consequence of the fact that
$\tr(\sigma^2)\leq 1$ for every state $\sigma$ on
$(\mathbb{C}^2)^{\otimes m}$, which implies $\sum_{\ntb\in\ncT}
\tr(\mysigma{\ntb}{}\sigma)^2\leq 2^m$ [cf.~\eqref{eq:sigmadecompos}].
Combining~\eqref{eq:derivatone}, ~\eqref{eq:dervtwo},
\eqref{eq:xyeentrop} with~\eqref{eq:Hbound} and using the fact that
$1+H(\nT) = H(\nK \nT)$ implies~\eqref{eq:goal} and thus concludes the
proof.
\end{proof}

Because of Lemma~\ref{lem:certainty}, we have $\Iacc(\nK \nT; \nE \nT)
= H(\nK \nT)$. Hence, together with Lemma~\ref{lem:mainlockinglemma}, we
conclude that the quantity $\Delta = \Iacc(\nK \nT;\nE \nT) -
\Iacc(\nK \nT;\nE)$, as defined in Section~\ref{sec:locking}, with
$\nV := (\nK, \nT)$, is arbitrarily close to $H(\nT) + 1$. We thus
have a locking effect: The difference $\Delta$ is larger than the size
of~$\nT$.

\section{Small accessible information does not imply secrecy} \label{sec:problem}

  
  
The locking property of the accessible information has dramatic
implications for cryptography. To illustrate this, we consider an
$n$-bit key $S = (S_1, \ldots, S_n)$ together with a quantum system
$\nE$ controlled by an adversary such that, for some bijective mapping
$f$, $\rho_{S_n f(S_1, \ldots, S_{n-1}) \nE} = \rho_{\nK \nT
  \nE}$~\footnote{That is, $\nK = S_n$ and $\nT = f(S_1, \ldots,
  S_{n-1})$.}, where $\rho_{\nK \nT \nE}$ is the \ccqstate{} as
defined in Section~\ref{sec:example} (for $m \approx n / \log_2
3$)~\footnote{As we will consider one-time pad encryption with the key
  $S$, we assume for simplicity that $S$ is a bitstring. Because the
  cardinality of the range of $(\nK, \nT)$ (i.e., $\{0,1\} \times
  \{1,2,3\}^m$) and $S$ (i.e., $\{0,1\}^n$) do not match, $S$ is not
  perfectly uniformly distributed on the key space.  However, a
  qualitatively identical statement with a perfectly uniformly
  distributed key can be obtained by using an appropriate adaption of
  the one-time pad to keys and messages on the space $\ncK \times \ncT
  = \{0,1\} \times \{1,2,3\}^m$. }.

It is an immediate consequence
of Lemma~\ref{lem:mainlockinglemma} that the key $S$ satisfies the
security criterion~\eqref{eq:Iacc} of Section~\ref{sec:intro}, i.e.,
\begin{equation} \label{eq:IaccSbound}
  \Iacc(S; \nE) = \Iacc(\nK \nT; \nE) \leq \eps \ ,
\end{equation}
where $\eps := e^{-\frac{n-2}{8}}$ decreases exponentially fast in the
key length $n$. However, as illustrated by the following example, this
is not sufficient for certain applications.

Assume that the key $S$ is used to encrypt an $n$-bit message $M=(M_1,
\ldots, M_n)$ by one-time pad encryption
and let $C=(C_1, \ldots, C_n)$ be the corresponding ciphertext.
Moreover, assume that an adversary has some a priori knowledge which
fully determines the first $n-1$ message bits $M_1, \ldots,
M_{n-1}$~\footnote{For example, the first $n-1$ bits of the message
  might be some redundant header information.}.  Upon receiving the
ciphertext bits $C_1, \ldots, C_{n-1}$, the adversary can thus easily
infer the first $n-1$ key bits $S_1, \ldots, S_{n-1}$. Hence, by
Lemma~\ref{lem:certainty}, she is now in a position to choose an
appropriate measurement of her quantum system $\nE$ which reveals the
$n$th key bit $S_n$ with certainty.  The encryption of the $n$th
message bit $M_n$ is thus completely insecure.



\section{Alternative security definition} \label{sec:alternative}

According to the discussion in the previous section, defining secrecy
with respect to the accessible information is problematic in a quantum
world. This raises the question whether there are stronger security
definitions which, e.g., imply that a secret key can safely be used
for one-time pad encryption. As shown
recently~\cite{BHLMO05,RenKoe05,Renner05}, the answer to this question
is positive~\footnote{It has been shown in~\cite{BHLMO05} that a key
  $S$ is secure in a strong sense if~\eqref{eq:IaccSbound} holds for a
  security parameter $\eps$ which is exponentially small in the key
  size. Our example shows that this exponential dependence is in fact
  necessary, thus answering an open question in~\cite{BHLMO05}.  Note,
  however, that making the security parameter $\eps$
  in~\eqref{eq:IaccSbound} exponentially small comes at the cost of
  reducing the key rate substantially.  }.

Let $\rho_{S E}$ be a \cqstate{} describing a classical key $S$
together with the quantum knowledge of an adversary, i.e., $\rho_{S E}
:= \sum_{s \in \cS} P_S(s) \proj{s} \otimes \rho_{E|S=s}$ where
$\{\ket{s}\}_{s \in \cS}$ are orthonormal states representing the
value of $S$.

\begin{definition}[\cite{RenKoe05,Renner05}] \label{def:sec}
  A random variable $S$ on $\cS$ is called an \emph{$\eps$-secure key
    with respect to $\nE$} if~\footnote{For two states $\rho$ and
    $\sigma$, $\|\rho - \sigma\| := 1/2 \tr |\rho-\sigma|$ denotes the
    \emph{trace distance} between $\rho$ and $\sigma$.}
\[
  \distb{\rho_{S E}}{\rho_U \otimes \rho_E} \leq \eps \ ,
\]
where $\rho_U := \sum_{s \in \cS} \frac{1}{|\cS|} \proj{s}$ is the
completely mixed state.
\end{definition}

As discussed in~\cite{RenKoe05}, $\eps$-security has an intuitive
interpretation: With probability $1-\eps$, the key $S$ can be
considered identical to a perfectly secure key $U$, i.e., $U$ is
uniformly distributed and independent of the adversary's information.
In other words, Definition~\ref{def:sec} guarantees that the key $S$
is perfectly secure except with probability $\eps$. Clearly, this is
still true if $S$ is used in any application.

Interestingly, this strong type of security can be achieved quite
easily. For example, it has been shown~\cite{RenKoe05} that the key
computed by applying a two-universal hash function to a random string
with sufficient entropy satisfies
Definition~\ref{def:sec}~\footnote{We refer to~\cite{RenKoe05}
  and~\cite{Renner05} for a detailed description of privacy
  amplification in the context of quantum adversaries.}. Security
proofs of QKD which are based on this result (see, e.g.,
\cite{KrGiRe05}) are thus not affected by the problem discussed above.


The following lemma shows that strongly secure keys can also be
obtained by measuring predistributed Bell states $\ket{\Phi^+}$ (or
approximations thereof). It follows from this statement that security
proofs based on entanglement purification (where the entanglement is
usually measured in terms of the fidelity to a fully entangled state,
as, e.g., in~\cite{LoCha99,ShoPre00}) can easily be adapted to meet
Definition~\ref{def:sec}~\footnote{These security proofs usually make
  use of a similar relation between the fidelity and the accessible
  information (see, e.g., Lemma~1 and~2 given in the supplementary
  material of~\cite{LoCha99} and the discussion in Footnote~28
  of~\cite{LoCha99}).  Substituting this relation by
  Lemma~\ref{lem:fid} thus turns these arguments into proofs of
  security according to the stronger criterion
  (Definition~\ref{def:sec}) given above.}  (see also \cite{BHLMO05}).

\begin{lemma} \label{lem:fid}
  Let $\eps \geq 0$ and let $\rho_{A B}$ be a bipartite quantum state
  such that $F(\rho_{AB}, \ket{\Phi^+}^{\otimes n}) \geq
  \sqrt{1-\eps^2}$. Then the two $n$-bit strings resulting from local
  measurements of $\rho_{A B}$ in the computational basis are
  $\eps$-secure keys (with respect to an adversary holding a
  purification of $\rho_{A B}$).
\end{lemma}

\begin{proof}
  According to Uhlmann's theorem, there exists a pure state
  $\ket{\kappa}$ and a purification $\ket{\Theta}$ of $\rho_{A B}$
  with some auxiliary system $\nE$ such that
  \[
    F(\ket{\Theta}, \ket{\Phi^+}^{\otimes n} \otimes \ket{\kappa}) 
  =
    F(\rho_{A B}, \ket{\Phi^+}^{\otimes n}) \ .
  \]
  Using the relation $\dist{\rho}{\sigma} \leq \sqrt{1-F(\rho,
    \sigma)^2}$ and the assumption of the lemma, we find
  \[
    \distb{\proj{\Theta}}{(\proj{\Phi^+} \otimes \proj{\kappa})^{\otimes n}} \leq \eps \ .
   \]
  Let $\rho_{S_A S_B E}$ be the \ccqstate{} describing the situation
   after measuring $\ket{\Theta}$ with respect to the computational
  basis in $A$ and $B$.  Because the trace distance can only decrease
  under physical operations, we conclude
  \[
    \distb{\rho_{S_A S_B E}}{\rho_{U U} \otimes \sigma_{E}} \leq \eps \ ,
  \]
  where $\rho_{U U} = \sum_{s \in \{0,1\}^n} \frac{1}{2^n} \proj{s}
  \otimes \proj{s}$.
\end{proof}

\section{Conclusions}

The setting considered in this paper consists of a classical $n$-bit
string $S=(S_1, \ldots, S_n)$ (for any $n \in \bbN$) and a quantum
system $\nE$ such that the following holds: (i)~any measurement on
$\nE$ chosen independently of $S$ only reveals a negligible amount of
information about $S$ (i.e., $\Iacc(S;\nE)$ is exponentially small in
$n$) and (ii)~given the first $n-1$ bits of $S$, there exists a
measurement on $\nE$ which determines the value of the $n$th bit with
certainty (i.e., $\Iacc(S; E S_1, \ldots, S_{n-1}) = n$).

This example of locking reveals a weakness of security definitions
based on the accessible information as they are used in the standard
literature on quantum cryptography. In particular, a secret key which
is secure according to such a definition might become completely
insecure when it is used in certain applications
(Section~\ref{sec:problem}). A possible solution to this problem is to
use the stronger yet still achievable notion of $\eps$-security
(Section~\ref{sec:alternative}): An $\eps$-secure key can safely be
used in any application---except with some (arbitrarily small)
probability~$\eps$.

\acknowledgments

Large parts of the work presented in this paper has been carried out
while the authors were with the Information Security and Cryptography
Research group of the Swiss Federal Institute of Technology (ETH)
Zurich, Switzerland. This research is partially funded by project
PROSECCO of the IST-FET programme of the EC and by Hewlett Packard.

\newcommand{\canc}[1]{#1}

\canc{

\appendix

\section*{Appendix}

Let $\rho_{\nV \nE}$ be a \cqstate{}. Lemma~\ref{lem:accesslower}
gives a lower bound on the entropy of $\nV$ conditioned on the outcome
of any measurement on $\nE$.

\begin{lemma}\label{lem:accesslower}
  Let $\rho_{\nV \nE}$ be a \cqstate{} with the property that
  $\rho_{\nE}$ is the completely mixed state on $\nE$. For some fixed
  POVM $\measp$ applied to $\nE$, let $H(\nV|\nZ)$ be the entropy of
  $\nV$ conditioned on the outcome $\nZ$. Then
  \[
    H(\nV|\nZ) \geq \min_{\sigma} H(\meas[\sigma])
  \]
  where $\sigma$ ranges over all states on $\nE$ and
  $H(\meas[\sigma])$ denotes the entropy of the outcome when the POVM
  $\meas := \{\dim(E) \cdot P_{\nV}(\nv)\cdot
  \rho_{\nE|\nV=\nv}\}_{\nv\in\ncV}$ is applied to $\sigma$.
\end{lemma}

\begin{proof}
  The fact that $\rho_{\nE}$ is the completely mixed state on $\nE$
  implies that $\meas$ is a POVM. The same fact also implies that the
  measurement result $\nZ$ is distributed according to
  $P_{\nZ}(\nz)=\frac{\tr(M_{\nz})}{d}$ for every outcome $\nz$, where
  $d := \dim(E)$ and $M_z$ are the operators of the POVM $\measp$.
  This in turn gives
\[
P_{\nV|\nZ}(\nv|\nz)=\frac{P_{\nV}(\nv)\cdot
    P_{\nZ|\nV}(\nz|\nv)}{P_{\nZ}(\nz)}=\frac{\tr(M_{\nz}\rho_{E|\nV=\nv})}{\tr(M_\nz)}\cdot
  d\cdot P_{\nV}(\nv)\ .
\]
Hence
\begin{align}\label{eq:lowerboundapplied}
H(\nV|\nZ)
\geq
\min_{\nz} H(\nV|\nZ=\nz)\ \geq \min_{\tilde{\sigma}}H(P^{\tilde{\sigma}}_{Z})\ ,
\end{align}
where the minimum is over all non-zero operators $\tilde{\sigma}$ on
$\nE$ with $0 \leq \tilde{\sigma}\leq \id_E$ and
$P_{\nV}^{\tilde{\sigma}}$ is the distribution
\[
P_{\nV}^{\tilde{\sigma}}(\nv):=\frac{\tr(\tilde{\sigma}\rho_{\nE|\nV=\nv})}{\tr(\tilde{\sigma})}\cdot
d\cdot P_{\nV}(\nv)\ .
\]
Note that for such an operator $\tilde{\sigma}$, the operator
$\sigma:=\frac{\tilde{\sigma}}{\tr(\tilde{\sigma})}$ is a state on
$\nE$. The assertion thus follows from~\eqref{eq:lowerboundapplied}
and the observation that $\meas[\sigma]\equiv
P_{\nV}^{\tilde{\sigma}}$.
\end{proof}

The next lemma summarises some properties of tensor products of Pauli
operators. As in Section~\ref{sec:example}, for any $m$-tuple $\ntb =
(\nt_1, \ldots, \nt_m)$ on $\{0, 1, 2, 3\}$, $\sigma_{\ntb}$ denotes the
$m$-fold tensor product $\sigma_{\nt_1} \otimes \cdots \otimes
\sigma_{\nt_m}$ of Pauli operators.

\begin{lemma}\label{lem:propertypauli}
  The following holds for all $m$-tuples $\ntb, \ntbp \in \{0,1,2,3\}^m$.
\begin{enumerate}[(i)]
\item\label{itemminus}
$\mysigma{\ntb}{}^\dagger=\mysigma{\ntb}{}$
\item\label{itemtwo}
$\tr(\mysigma{\ntb}{})= 2^m \cdot \delta_{\ntb, 0}$.
\item\label{itemthree} The eigenvalues of $\mysigma{\ntb}{}$ are
  $\{-1,1\}$.
\item\label{itemzero} $\tr(\mysigma{\ntb}{}^\dagger
  \mysigma{\ntbp}{})=2^m \cdot \delta_{\ntb, \ntbp}$.
\end{enumerate}
\end{lemma}

Lemma~\ref{lem:propertypauli} implies that the operators
$\{2^{-\frac{m}{2}}\cdot \mysigma{\ntb}{}\}_{\ntb \in \{0, 1,2,3\}^m}$
form an orthonormal basis of the space of hermitian operators on
$(\mathbb{C}^2)^{\otimes m}$ with respect to the Hilbert-Schmidt
scalar product $\langle A,B\rangle:=\tr(A^\dagger B)$. In particular,
every state $\sigma$ on $(\mathbb{C}^2)^{\otimes m}$ can be written in
the so-called generalised Bloch representation as
\begin{align}\label{eq:sigmadecompos}
\sigma=2^{-m}\sum_{\ntb\in\{0, 1,2,3\}^m}
\tr(\mysigma{\ntb}{}\sigma)\mysigma{\ntb}{}\ ,
\end{align}
where the coefficients $\tr(\mysigma{\ntb}{}\sigma)$ are real-valued.

}


\begin{thebibliography}{17}
\expandafter\ifx\csname natexlab\endcsname\relax\def\natexlab#1{#1}\fi
\expandafter\ifx\csname bibnamefont\endcsname\relax
  \def\bibnamefont#1{#1}\fi
\expandafter\ifx\csname bibfnamefont\endcsname\relax
  \def\bibfnamefont#1{#1}\fi
\expandafter\ifx\csname citenamefont\endcsname\relax
  \def\citenamefont#1{#1}\fi
\expandafter\ifx\csname url\endcsname\relax
  \def\url#1{\texttt{#1}}\fi
\expandafter\ifx\csname urlprefix\endcsname\relax\def\urlprefix{URL }\fi
\providecommand{\bibinfo}[2]{#2}
\providecommand{\eprint}[2][]{\url{#2}}

\bibitem[{\citenamefont{DiVincenzo et~al.}(2004)\citenamefont{DiVincenzo,
  Horodecki, Leung, Smolin, and Terhal}}]{DHLST04}
\bibinfo{author}{\bibfnamefont{D.~P.} \bibnamefont{DiVincenzo}},
  \bibinfo{author}{\bibfnamefont{M.}~\bibnamefont{Horodecki}},
  \bibinfo{author}{\bibfnamefont{D.~W.} \bibnamefont{Leung}},
  \bibinfo{author}{\bibfnamefont{J.~A.} \bibnamefont{Smolin}},
  \bibnamefont{and} \bibinfo{author}{\bibfnamefont{B.~M.}
  \bibnamefont{Terhal}}, \bibinfo{journal}{Phys.\ Rev.\ Lett.}
  \textbf{\bibinfo{volume}{92}}, \bibinfo{pages}{067902}
  (\bibinfo{year}{2004}).

\bibitem[{\citenamefont{Buhrman et~al.}(2005)\citenamefont{Buhrman, Christandl,
  Hayden, Lo, and Wehner}}]{BCHLW05}
\bibinfo{author}{\bibfnamefont{H.}~\bibnamefont{Buhrman}},
  \bibinfo{author}{\bibfnamefont{M.}~\bibnamefont{Christandl}},
  \bibinfo{author}{\bibfnamefont{P.}~\bibnamefont{Hayden}},
  \bibinfo{author}{\bibfnamefont{H.-W.} \bibnamefont{Lo}}, \bibnamefont{and}
  \bibinfo{author}{\bibfnamefont{S.}~\bibnamefont{Wehner}}
  (\bibinfo{year}{2005}), \bibinfo{note}{eprint: quant-ph/0504078}.

\bibitem[{\citenamefont{Hayden et~al.}(2004)\citenamefont{Hayden, Leung, Shor,
  and Winter}}]{HLSW04}
\bibinfo{author}{\bibfnamefont{P.}~\bibnamefont{Hayden}},
  \bibinfo{author}{\bibfnamefont{D.}~\bibnamefont{Leung}},
  \bibinfo{author}{\bibfnamefont{P.~W.} \bibnamefont{Shor}}, \bibnamefont{and}
  \bibinfo{author}{\bibfnamefont{A.}~\bibnamefont{Winter}},
  \bibinfo{journal}{Comm.\ Math.\ Phys.} \textbf{\bibinfo{volume}{250}},
  \bibinfo{pages}{371} (\bibinfo{year}{2004}).

\bibitem[{\citenamefont{Ben-Or et~al.}(2005)\citenamefont{Ben-Or, Horodecki,
  Leung, Mayers, and Oppenheim}}]{BHLMO05}
\bibinfo{author}{\bibfnamefont{M.}~\bibnamefont{Ben-Or}},
  \bibinfo{author}{\bibfnamefont{M.}~\bibnamefont{Horodecki}},
  \bibinfo{author}{\bibfnamefont{D.~W.} \bibnamefont{Leung}},
  \bibinfo{author}{\bibfnamefont{D.}~\bibnamefont{Mayers}}, \bibnamefont{and}
  \bibinfo{author}{\bibfnamefont{J.}~\bibnamefont{Oppenheim}}, in
  \emph{\bibinfo{booktitle}{Second Theory of Cryptography Conference, {TCC}
  2005}} (\bibinfo{year}{2005}), vol. \bibinfo{volume}{3378}, pp.
  \bibinfo{pages}{386--406}.

\bibitem[{\citenamefont{Renner and K\"onig}(2005)}]{RenKoe05}
\bibinfo{author}{\bibfnamefont{R.}~\bibnamefont{Renner}} \bibnamefont{and}
  \bibinfo{author}{\bibfnamefont{R.}~\bibnamefont{K\"onig}}, in
  \emph{\bibinfo{booktitle}{Second Theory of Cryptography Conference, {TCC}
  2005}} (\bibinfo{year}{2005}), vol. \bibinfo{volume}{3378}, pp.
  \bibinfo{pages}{407--425}.

\bibitem[{\citenamefont{Renner}(2005)}]{Renner05}
\bibinfo{author}{\bibfnamefont{R.}~\bibnamefont{Renner}}, Ph.D. thesis,
  \bibinfo{school}{ETH Z\"urich} (\bibinfo{year}{2005}).

\bibitem[{\citenamefont{Kraus et~al.}(2005)\citenamefont{Kraus, Gisin, and
  Renner}}]{KrGiRe05}
\bibinfo{author}{\bibfnamefont{B.}~\bibnamefont{Kraus}},
  \bibinfo{author}{\bibfnamefont{N.}~\bibnamefont{Gisin}}, \bibnamefont{and}
  \bibinfo{author}{\bibfnamefont{R.}~\bibnamefont{Renner}},
  \bibinfo{journal}{Phys.\ Rev.\ Lett.} \textbf{\bibinfo{volume}{95}},
  \bibinfo{pages}{080501} (\bibinfo{year}{2005}).

\bibitem[{\citenamefont{Lo and Chau}(1999)}]{LoCha99}
\bibinfo{author}{\bibfnamefont{H.-K.} \bibnamefont{Lo}} \bibnamefont{and}
  \bibinfo{author}{\bibfnamefont{H.~F.} \bibnamefont{Chau}},
  \bibinfo{journal}{Science} \textbf{\bibinfo{volume}{283}},
  \bibinfo{pages}{2050} (\bibinfo{year}{1999}).

\bibitem[{\citenamefont{Shor and Preskill}(2000)}]{ShoPre00}
\bibinfo{author}{\bibfnamefont{P.}~\bibnamefont{Shor}} \bibnamefont{and}
  \bibinfo{author}{\bibfnamefont{J.}~\bibnamefont{Preskill}},
  \bibinfo{journal}{Phys.\ Rev.\ Lett.} \textbf{\bibinfo{volume}{85}}
  (\bibinfo{year}{2000}).

\bibitem[{\citenamefont{Vernam}(1926)}]{Vernam26}
\bibinfo{author}{\bibfnamefont{G.~S.} \bibnamefont{Vernam}},
  \bibinfo{journal}{J. Am. Inst. Elec. Eng.} \textbf{\bibinfo{volume}{55}}
  (\bibinfo{year}{1926}).

\bibitem[{\citenamefont{Maurer}(1993)}]{Maurer93}
\bibinfo{author}{\bibfnamefont{U.~M.} \bibnamefont{Maurer}},
  \bibinfo{journal}{IEEE Transactions on Information Theory}
  \textbf{\bibinfo{volume}{39}}, \bibinfo{pages}{733} (\bibinfo{year}{1993}).

\bibitem[{\citenamefont{Wyner}(1975)}]{Wyner75}
\bibinfo{author}{\bibfnamefont{A.~D.} \bibnamefont{Wyner}},
  \bibinfo{journal}{Bell System Technical Journal}
  \textbf{\bibinfo{volume}{54}} (\bibinfo{year}{1975}).

\bibitem[{\citenamefont{Bennett et~al.}(1988)\citenamefont{Bennett, Brassard,
  and Robert}}]{BeBrRo88}
\bibinfo{author}{\bibfnamefont{C.~H.} \bibnamefont{Bennett}},
  \bibinfo{author}{\bibfnamefont{G.}~\bibnamefont{Brassard}}, \bibnamefont{and}
  \bibinfo{author}{\bibfnamefont{J.-M.} \bibnamefont{Robert}},
  \bibinfo{journal}{SIAM Journal on Computing} \textbf{\bibinfo{volume}{17}},
  \bibinfo{pages}{210} (\bibinfo{year}{1988}).

\bibitem[{\citenamefont{Csisz\'{a}r and K\"{o}rner}(1978)}]{CsiKor78}
\bibinfo{author}{\bibfnamefont{I.}~\bibnamefont{Csisz\'{a}r}} \bibnamefont{and}
  \bibinfo{author}{\bibfnamefont{J.}~\bibnamefont{K\"{o}rner}},
  \bibinfo{journal}{IEEE Transactions on Information Theory}
  \textbf{\bibinfo{volume}{24}}, \bibinfo{pages}{339} (\bibinfo{year}{1978}).

\bibitem[{\citenamefont{Nielsen and Chuang}(2000)}]{NieChu00}
\bibinfo{author}{\bibfnamefont{M.~A.} \bibnamefont{Nielsen}} \bibnamefont{and}
  \bibinfo{author}{\bibfnamefont{I.~L.} \bibnamefont{Chuang}},
  \emph{\bibinfo{title}{Quantum computation and quantum information}}
  (\bibinfo{publisher}{Cambridge University Press}, \bibinfo{year}{2000}).

\bibitem[{\citenamefont{Gottesman and Lo}(2003)}]{GotLo03}
\bibinfo{author}{\bibfnamefont{D.}~\bibnamefont{Gottesman}} \bibnamefont{and}
  \bibinfo{author}{\bibfnamefont{H.-K.} \bibnamefont{Lo}},
  \bibinfo{journal}{IEEE Transactions on Information Theory}
  \textbf{\bibinfo{volume}{49}}, \bibinfo{pages}{457} (\bibinfo{year}{2003}).

\bibitem[{\citenamefont{Lo et~al.}(2005)\citenamefont{Lo, Chau, and
  Ardehali}}]{LoChAr05}
\bibinfo{author}{\bibfnamefont{H.-K.} \bibnamefont{Lo}},
  \bibinfo{author}{\bibfnamefont{H.~F.} \bibnamefont{Chau}}, \bibnamefont{and}
  \bibinfo{author}{\bibfnamefont{M.}~\bibnamefont{Ardehali}},
  \bibinfo{journal}{Journal of Cryptology} \textbf{\bibinfo{volume}{18}},
  \bibinfo{pages}{133} (\bibinfo{year}{2005}).

\end{thebibliography}


\end{document}